%% file: sample_saj.tex
\documentclass{article_saj}
\usepackage{graphicx,saj,multicol,subeqnarray}
\usepackage{float}
\usepackage{xcolor}
\usepackage{widetext}
\usepackage{url}
\usepackage{subfigure} 
\usepackage{bm}
\usepackage{tikz} 
\usepackage{pifont} 
\usepackage{amsfonts}
\usepackage{amssymb}
\usepackage{amsmath,upgreek}
\usepackage{titlesec}

\input{astro}

\titlelabel{\thetitle.\quad}
\definecolor{xlinkcolor}{cmyk}{1,0.6,0,0}
\usepackage[bookmarks=false,         
     pdfnewwindow=true,      
     colorlinks=true,    
     linkcolor=xlinkcolor,     
     citecolor=xlinkcolor,     
     filecolor=xlinkcolor,  
     urlcolor=xlinkcolor,      
final=true
]{hyperref}

\def\papertype{\ \hfill\ Editorial}

\def\udc{52}
\begin{document}
\parindent=.5cm
\baselineskip=3.8truemm
\columnsep=.5truecm
\newenvironment{lefteqnarray}{\arraycolsep=0pt\begin{eqnarray}}
{\end{eqnarray}\protect\aftergroup\ignorespaces}
\newenvironment{lefteqnarray*}{\arraycolsep=0pt\begin{eqnarray*}}
{\end{eqnarray*}\protect\aftergroup\ignorespaces}
\newenvironment{leftsubeqnarray}{\arraycolsep=0pt\begin{subeqnarray}}
{\end{subeqnarray}\protect\aftergroup\ignorespaces}
%


\markboth{\eightrm Nonlinear identification algorithm for online and offline study of pulmonary mechanical ventilation} 
{\eightrm D. A. Riva {\lowercase{\eightit{et al.}}}}

\begin{strip}

{\ }

\vskip-1cm



{\ }


\title{Nonlinear identification algorithm for online and offline study of pulmonary mechanical ventilation}


\authors{Diego A. Riva$^{1}$,  Carolina A. Evangelista$^{1}$, Paul F. Puleston$^{1}$, Luis Corsiglia$^2$ and Nahuel Dargains$^2$}

\vskip3mm


\address{$^1$Instituto LEICI, Facultad de Ingeniería, Universidad Nacional de La Plata-CONICET, AR}


\Email{diego.riva@ing.unlp.edu.ar}

\address{$^2$Hospital I.E.A y C. San Juan de Dios de La Plata, AR }





\summary{This work presents an algorithm for determining the parameters of a nonlinear dynamic model of the respiratory system in patients undergoing assisted ventilation. Using the pressure and flow signals measured at the mouth, the model's quadratic pressure-volume characteristic is fit to this data in each respiratory cycle by appropriate estimates of the model parameters. Parameter changes during ventilation can thus also be detected. The algorithm is first refined and assessed using data derived from simulated patients represented through a sigmoidal pressure-volume characteristic with hysteresis. As satisfactory results are achieved with the simulated data, the algorithm is evaluated with real data obtained from actual patients undergoing assisted ventilation. The proposed nonlinear dynamic model and associated parameter estimation algorithm yield closer fits than the static linear models computed by respiratory machines, with only a minor increase in computation. They also provide more information to the physician, such as the pressure-volume (P-V) curvature and the condition of the lung (whether normal, under-inflated, or over-inflated). This information can be used to provide safer ventilation for patients, for instance by ventilating them in the linear region of the respiratory system.}


\keywords{Mechanical Ventilation -- Nonlinear Identification -- Dynamic Pulmonary Models -- 
Diagnosis Tool}

\end{strip}

\tenrm


\section{INTRODUCTION}

\indent

\section{Introduction}
From the point of view of intensive medicine, mechanical ventilation therapy is a procedure used to maintain breathing temporarily, for the time necessary until the patient recovers their functional capacity and can resume spontaneous ventilation \cite{gattinoni2017future}. The ventilatory strategies are based on calculated parameters and help patient monitoring against adverse effects of mechanical ventilation \cite{grasselli2021mechanical}, \cite{silva2018basics}. 

A patient connected to a breathing machine offers the opportunity to compute various physiological parameters and to evaluate their modifications to changes in the breathing support strategy \cite{borges2013effects}. However, ventilators in intensive care units usually provide to the clinicians a rough estimation of two key physiological parameters: the airway resistance and the compliance of the breathing system \cite{walter2018invasive}. Standard equipment typically computes them considering no more than five points of the total data available from the respiratory cycles: the Plateau Pressure, the Peak Inspiratory Pressure (PIP), the Peak Inspiratory Flow (PIF), the Total Inspiration Volume, and the Positive End-Expiratory Pressure (PEEP) \cite{baconnier1995computer}. A few devices make use of the whole cycle data, such as Hamilton ventilators, which obtain values for the two parameters through least squares fitting based on a linear model of the system \cite{iotti1999measurements}.

In this sense, model-based methods which use all the data from the cycle measured by the ventilator fit the respiratory behaviour better, giving clinicians a deeper physiological picture of the patient’s state \cite{schranz2011hierarchical}, \cite{redmond2019evaluation}, \cite{damanhuri2016assessing}, \cite{chase2018next}, \cite{morton2019optimising}. Such models can be used to optimise the ventilator settings, i.e., to reduce harmful effects of the ventilation therapy or to keep the patient within a ``zone of comfort" during mechanical ventilation \cite{kim2020model}. Nevertheless, most of those models characterise the dynamics of the pulmonary system as linear, simplifying their description at the expense of accuracy \cite{schranz2012iterative}, \cite{lucangelo2005respiratory}.

In this paper, a more accurate nonlinear model is considered to better represent the respiratory system of a sedated patient under assisted ventilation, especially when there is pulmonary disease, as the physiology of their lungs may change remarkably and cause them to breath in the nonlinear regions \cite{grinnan2005clinical}. Thus, an algorithm based on nonlinear identification techniques is proposed to estimate the model parameters, which uses a hierarchical identification process. Firstly, the parameters of a linear version of the model are estimated and then, those parameters are used for the initialisation and identification of the nonlinear model. In this way, with the flow and the pressure data measured in the patient's mouth, the algorithm computes a set of parameters of the nonlinear model for each respiratory cycle.

This work comprises two stages. In the first one, a simulated patient is initially used to assess the algorithm efficiency, considering nonlinear sigmoid models. Then, an even more realistic nonlinear model which includes the occurrence of local hystereses is implemented, to evaluate the capability of the algorithm to get information from more realistic data. After the successful in-silico validation, in the second stage the proposed algorithm is applied to real respiratory systems of sedated patients with a breathing disease, particularly COVID-19. To this end, measurements of flow and pressure obtained from PEEP titration manoeuvres are utilised, since during these manoeuvres the patient is ventilated in a wide range of breathing regions.

The main contributions of this work are listed as follows:
\begin{itemize}
    \item An algorithm that carries out an identification process to obtain the nonlinear model parameters of a patient under assisted ventilation is developed.
    \item The algorithm can determine the respiratory region where the patient is ventilated, i.e. Atelectasis, Linear, and Overdistension Regions.
    \item The proposed parameter identification algorithm is validated on extensive simulated data (including noise corrupted data), which uses a sigmoidal with hysteresis behaviour to describe patient compliance.
    \item The algorithm is tested using actual patients' data, and the results confirm its potential value in critical care.
\end{itemize}

The remainder of this article is organised as follows. Section \ref{sec:Respiratory physiology} introduces respiratory physiology to understand the mathematical models used in this work, which are explained in Section \ref{sec:models}. Section \ref{sec:NonlinearIDalgorithm} presents the developed algorithm used to obtain the parameter values of the models. Section \ref{sec:Results} presents the results obtained using simulated and actual patients under assisted ventilation. Finally, Section \ref{sec:Conclusions} concludes the paper, highlighting the main findings and describing future research directions.

\section{Respiratory physiology} \label{sec:Respiratory physiology}

To understand the models that describe the behaviour of the pulmonary system, some respiratory physiology concepts are explained in this section.

Air moves due to a pressure gradient. If external pressure (pressure in the mouth) is higher than internal pressure (pressure in respiratory system), entry of air into the pulmonary system occurs (inhalation). By contrast, if external pressure is less than the internal, the system expels air (exhalation). These changes in pressure can occur spontaneously (carried out by respiratory muscles) or artificially (performed by a mechanical ventilator).

\subsection{Spontaneous ventilation}

Considering atmospheric pressure as the base reference level, the respiratory system achieves the entry of air from variations of the internal pressure, generated by the respiratory muscles. In spontaneous ventilation, this pressure variation is responsible for the instantaneous value of the air volume moving in or out of the lungs during the respiratory process.

\begin{figure}[ht]
    \centering
    \includegraphics[width=\columnwidth]{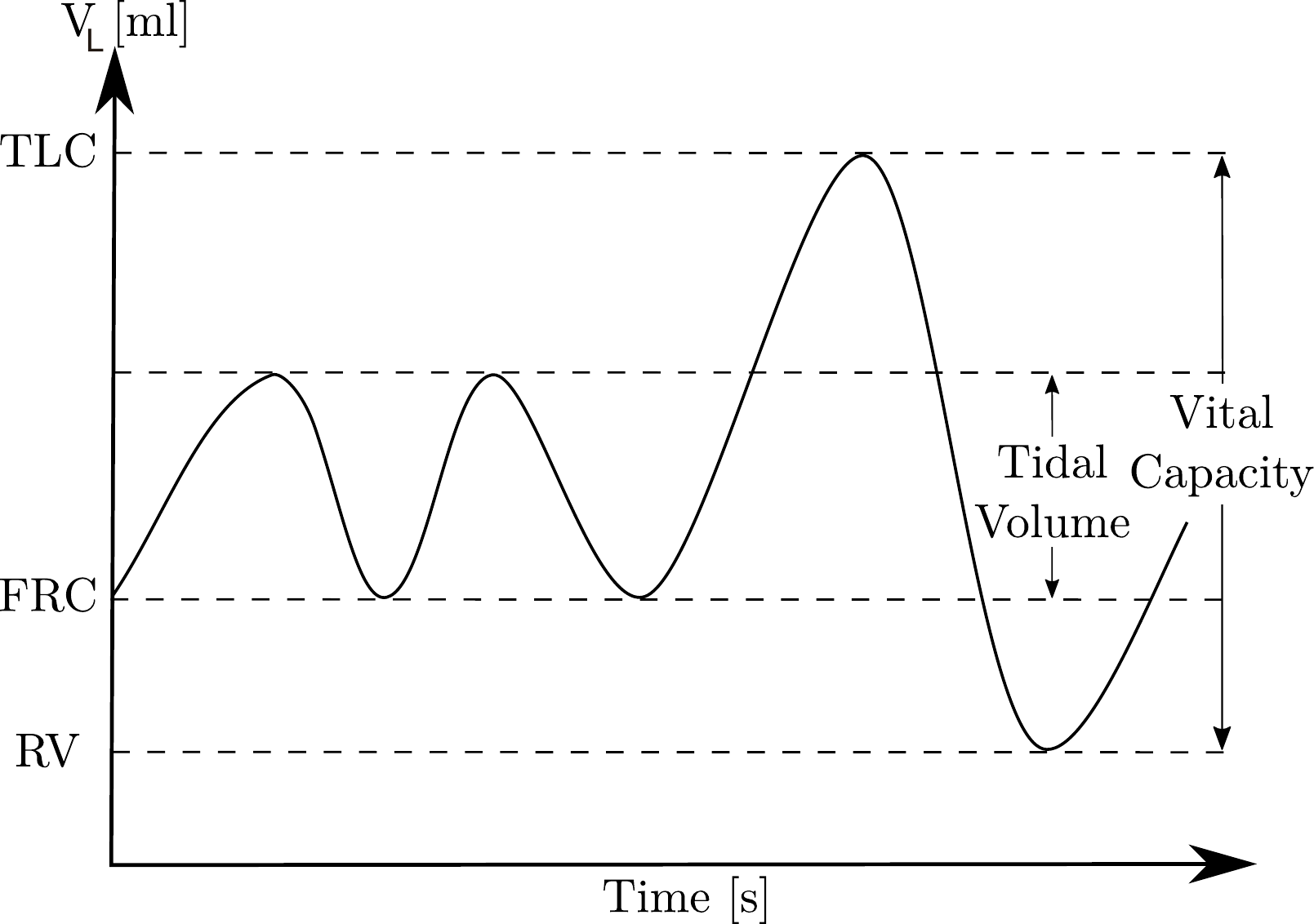}
    \caption{\label{fig:Volumenes} Characterisation of volumes in the system. 
    }
\end{figure}

Defining the instantaneous volume of the lung at time t as $V_L(t)$, it can be observed in Fig. \ref{fig:Volumenes} that the total amount of air that enters with each breath (known as \textit{Tidal Volume} $V_T$) is considerably less than the total volume that can be forcibly inhaled from a maximal inspiration following a maximal exhalation, called \textit{Vital Capacity}.
Defining the \textit{Residual Volume} (RV) as the amount of air that remains in a person's lungs after a maximum exhalation, the \textit{Total Lung Capacity} (TLC) is the total lung volume, considering the Vital Capacity plus the Residual Volume, as it is schematised in Fig. \ref{fig:Volumenes} \cite{west2012respiratory}. Finally, after a normal passive exhalation, the amount of air remaining in the lungs is called \textit{Functional Residual Capacity} (FRC), which is a volume greater than the Residual Volume. The term $V(t)$ is used to denote $V_L(t)-FRC$.

\subsection{Mechanical ventilation} \label{ssec:ventilation}

In the event that the patient is unable to breathe on their own, either totally or partially, mechanical ventilation can be used to maintaining their good oxygenation and monitoring their lung mechanics.

The pressure imposed by the ventilator at the patient’s mouth, $P(t)$, is composed by a variable component, $P_v(t)$,  mounted on a programmable constant pressure, the PEEP (i.e., \mbox{$P(t)=P_v(t)+PEEP$)}. $P_v(t)$ establishes the aforementioned variable volume $V(t)$, ensuring a predefined respiratory cycle set by the clinician. On the other hand, the PEEP is a ventilator parameter, also configurable by the clinician, used to attain that at the end of expiration the lung does not return to atmospheric pressure, hence preventing the collapse of alveoli every cycle \cite{sahetya2020peep} \cite{van2019recruitment}. The PEEP, adequately set, can decrease the respiratory work and improve the patient's oxygenation, since it may increase the FRC of the patient \cite{dries2020finding}.

Most of the ventilation machines can work in one of the following two types of operation. In the first one, assisted ventilation, the machine assists the patient after the detection of an inspiratory effort and so, the patient can control the timing and size of the breath. In the second type, controlled ventilation, a patient effort is not required, and the ventilator starts automatically every respiratory cycle with a predefined period \cite{tobin1994mechanical}. When in this operation type, the patient needs to be sedated to avoid asynchronies.

The breathing pattern produced by the ventilator requires the setting of control parameters such as the size of the breath, how fast and how often air is brought in and let out, and how much effort, if any, the patient must exert to signal the ventilator to start a breath. Numerous `ventilation modes' have been developed to make ventilators produce breathing patterns that coordinate the machine's activity with the needs of the patient \cite{tobin1994mechanical}. The best known are:

\begin{itemize}
    \item \textbf{Volume control ventilation (VCV)}. In this ventilation mode, the physician sets the volume of air that will enter the patient in each respiratory cycle. The ventilator will then establish a flow profile for inspiration, and the pressure in the mouth will depend on the patient's respiratory mechanics.
    \item \textbf{Pressure control ventilation (PCV)}. The physician sets the peak value of pressure that the ventilator will reach in each respiration cycle. In this case, the amount of air volume entering the lungs is not fixed, but instead depends on the respiratory mechanics of the corresponding patient. 
\end{itemize}

\subsection{Hysteresis and Pressure-Volume curves}\label{sec:hysteresis}

\begin{figure}[ht]
    \centering
    \includegraphics[width=0.5\textwidth]{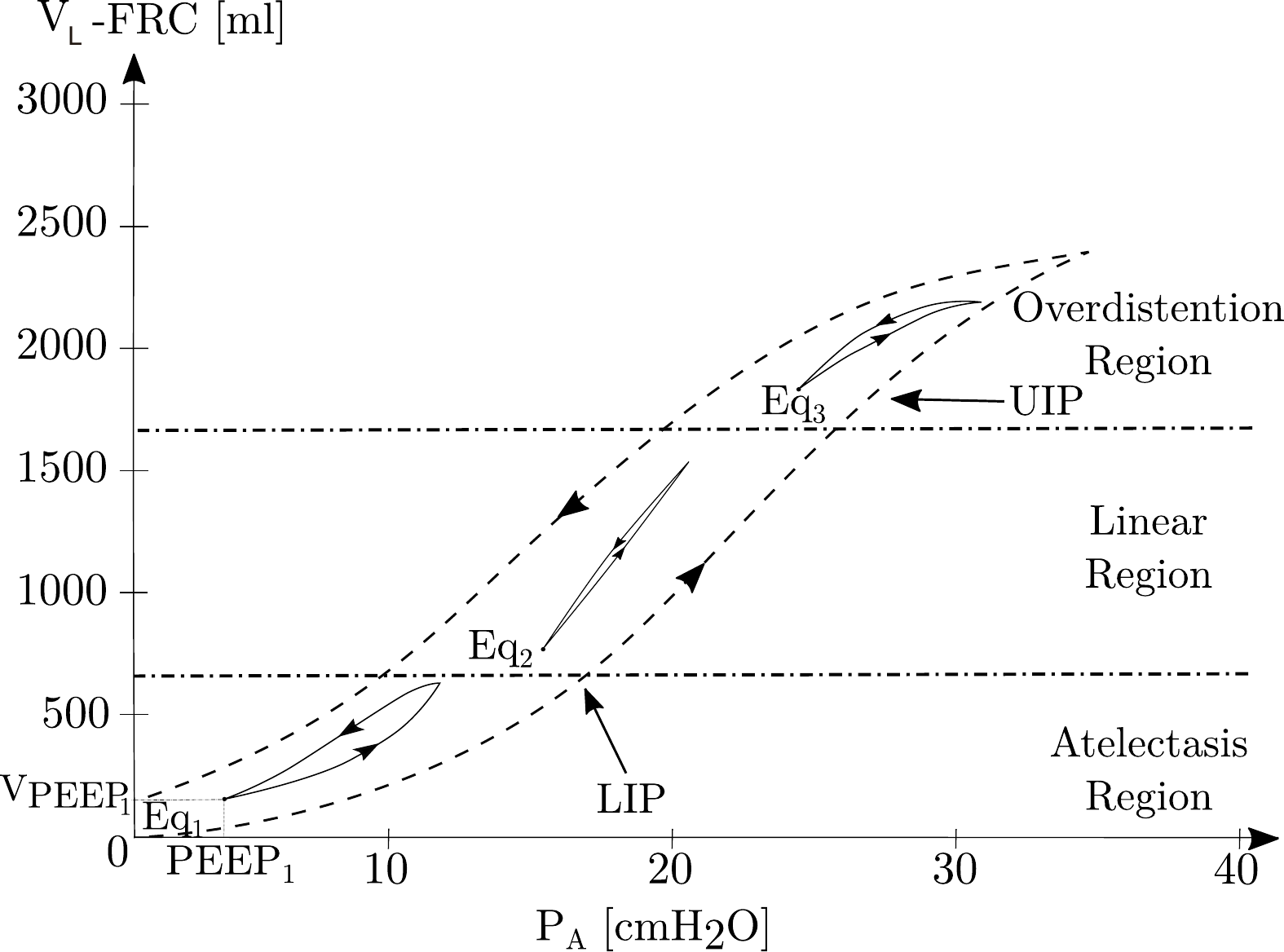}
    \caption{\label{fig:HisteresisCurvePV} In dashed lines, the Q-S P-V curve measured from FRC to TLC is shown. In solid lines, illustrative respiratory cycles curves starting from different equilibrium points $Eq_i$, \mbox{$i=1,2,3$}, showing the various curve shapes that can be observed depending on the region where the patient is ventilated. $P_A$ is the alveolar pressure, $V_{L}$ is the total volume and $V_{PEEP_i}$ is the volume due to the ventilator $PEEP_i$.} 
\end{figure}

The Quasi-Static Pressure-Volume (Q-S P-V) curve represents the limits of pressure and volume inside of which respiratory cycle paths can take place \cite{rimensberger1999lung} (see Fig. \ref{fig:HisteresisCurvePV}). It covers a range from FRC to TLC and its contour can be empirically obtained by measuring pressure in the mouth while maintaining lengthy low and constant inspiration and expiration flows, so that the pressure drops in the airways are negligible. Under these conditions, $P(t)$ equals the alveolar pressure $P_A$. \cite{harris2005pressure} \cite{lu1999simple}.

A physiological phenomenon that can be observed in a patient is the hysteresis. Hysteresis is defined as different volume paths upon application and withdrawal of a certain pressure, and reflects an imperfect elastic response to deformation \cite{escolar2004lung} \cite{fedullo1980hysteresis}.

Additionally, two characteristic inflection points can be appreciated in the curve, denoted as LIP (i.e. Low Inflection Point) and UIP (i.e. Upper Inflection Point) \cite{venegas1998comprehensive}. They define the three main respiratory regions of the patient: (a) the Atelectasis Region: from the LIP zone downwards, the lung is poorly recruited and many alveoli must be opening and closing with each respiratory cycle \cite{tsuchida2006atelectasis}; (b) the Linear Region: between the LIP and the UIP, is the desired breathing region where alveoli are highly recruited plus their elasticity is at peak performance; and (c) the Overdistension Region: from the UIP zone upwards, the lung is functioning near its elastic limit and overdistension of some alveoli units can take place, causing injury to the patient \cite{gonzalez2012lung}.

\begin{figure}[ht]
    \centering
    \includegraphics[width=0.45\textwidth]{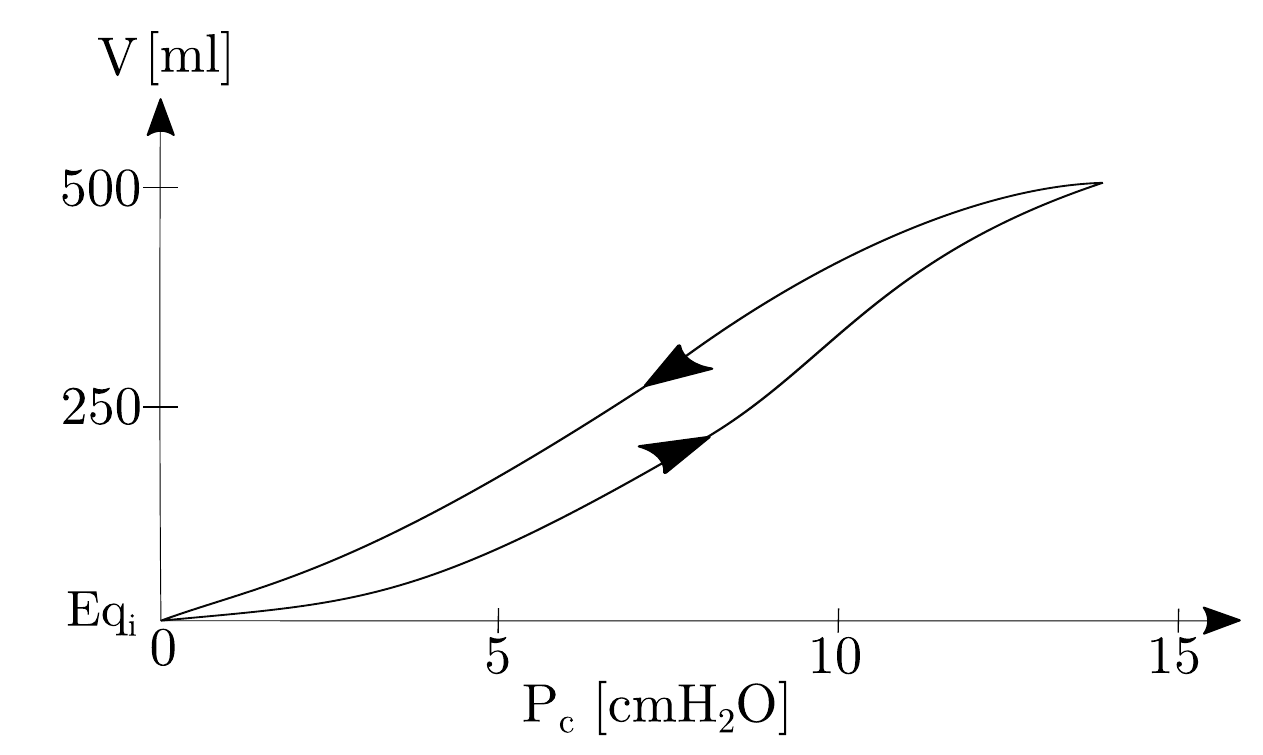}
   \caption{\label{fig:HisteresisLocal} Hysteresis in the lungs-chest wall system in a typical respiratory cycle of a  patient. The arrow pointing upwards corresponds to the inspiration path, and the arrow pointing downwards to the expiration path.}
\end{figure}

Some illustrative respiratory hysteresis cycles are depicted in each one of the respiratory regions in Fig. \ref{fig:HisteresisCurvePV}. Such respiratory cycles start on different initial equilibrium points ($Eq_i$, $i=1,2,3$ in Fig. \ref{fig:HisteresisCurvePV}), which depend on the patient's FRC plus the extra air volume due to the ventilator PEEP \cite{muniz2009pressure}. From a given equilibrium, 
Volume $V(t)$ evolves in accordance with the variable component ($P_v(t)$) of the ventilator pressure. A diagram of a generic respiratory cycle hysteresis which starts from an $Eq_i$ is shown in Fig. \ref{fig:HisteresisLocal}.

\section{Respiratory system modelling}\label{sec:models}

In this section the equation of motion that relates pressures in the respiratory system with the volume of air moving in and out of lungs-chest wall system is presented. Subsequently, the corresponding linear and nonlinear models are introduced, whose parameters will be estimated by the proposed identification algorithm.
 
\subsection{The equation of motion}

The breathing dynamics of a sedated patient can be described by relating the pressure applied in the patient's mouth with the volume and flow in the respiratory system.

\begin{figure}[ht]
 

\centering
\includegraphics[width=0.27\textwidth]{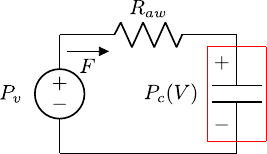}
\caption{\label{fig:ModElec} Electrical equivalent model of the respiratory system.}
\end{figure}

The elastic properties are determined by the relationship between lung volume and pressure in the lungs-chest wall system. The resistive properties are represented by the airway resistance. Such relations, starting from a given $Eq_i$, are shown in Eq. \eqref{eq:Eq_Motion}, commonly known as Equation of Motion \cite{bates2009lung}, and its electrical equivalent model is represented in Fig. \ref{fig:ModElec}:
\begin{equation}\label{eq:Eq_Motion}
    P_v(t)=\dot{V}(t)R_{aw}+P_{c}(V(t))
\end{equation}
where the time derivative $\dot{V}(t)$ of $V(t)$ corresponds to the airflow $F(t)$ into the lungs. $R_{aw}$ represents the airflow resistance of the airways and $P_{c}(V(t))$ is the variable pressure in the lungs-chest wall system, which depends on $V(t)$ and on the elastic properties of the region where the patient is being ventilated. To simplify notation, the time dependency will not be written in the following.

In the electrical equivalent model shown in Fig. \ref{fig:ModElec}, the capacitor characterised by $P_{c}(V)$ corresponds to the lung-chest wall compliance.

A widely accepted characterisation of the lung-chest wall system in the complete volume range from the FRC to the TCL corresponds to a sigmoid curve \cite{harris2005pressure}. However, when describing normal either healthy or unhealthy breathing cycles, simpler `local' descriptions can be used, as the sigmoid curve can be well approximated via piecewise continuous polynomial functions depending on the $V$ and the respiratory region.

A linear model for $P_{c}(V)$ provides a very accurate approximation in the Linear Region, but a coarse one outside of it. Thus, in this work, a quadratic (nonlinear) model for $P_{c}(V)$ is used for identification, to be able to obtain an accurate approximation also in the Atelectasis and Overdistension Regions. Both the Linear and Nonlinear identification models are presented in sections \ref{ssec:modelsLM}, and \ref{ssec:modelsNLM}, respectively.

\subsection{Respiratory System Linear Model (LM)}\label{ssec:modelsLM}

The brief physiological explanation of the respiratory system in subsection \ref{sec:hysteresis} helps understanding the complexity of accurately modelling its dynamics. However, there are simple models that still provide the most significant information. In that sense, the linear model is the most widespread model among clinicians. According to it, the relationship between $P_{c}$ and $V$ is characterised as:
\begin{equation}\label{eq:Lineal}
    P_{c}(V)=\frac{V}{C}=VE
\end{equation}
where $C$, the compliance of the respiratory system, is a constant which depends on the region of the P-V curve where breathing is taking place and, of course, on the health condition of the patient. The inverse of the compliance is known as elastance ($E=1/C$). The value of $C$ is maximum when the patient is ventilated in the Linear Region.

Thus, using the linear description of Eq. \eqref{eq:Lineal} and the equation of motion, Eq. \eqref{eq:Eq_Motion}, the dynamics of the linear model can be expressed as:
\begin{equation}
     \dot{V}= -\frac{1}{R_{aw} C} V + \frac{P_v}{R_{aw}}
\end{equation}
It can be observed that two parameters characterise the patient's respiratory system in the region where breathing is occurring:
\begin{equation}\label{eq:linear_parameter}
    \theta_l=[C, R_{aw}]
\end{equation}
They will be estimated via the identification algorithm developed and presented in this work, by using the airflow $F=\dot{V}$ and pressure $P_v$ measured on the patient.

\subsection{Respiratory System Nonlinear Model (NLM)}\label{ssec:modelsNLM}

A better characterisation of the relation between $P_{c}$ and $V$ can be obtained by using a quadratic description:
\begin{equation}\label{eq:noLineal}
    P_{c}(V)=V(a_1 + a_2V)
\end{equation}
where the elastance varies with $V$. Eq. \eqref{eq:noLineal} provides improved fit over that of the linear model, particularly when the lungs are ventilated in the areas near the inflection points (UIP or LIP) \cite{bates2009lung}.  Additionally, the concavity of the curve obtained in Eq. \eqref{eq:noLineal} can provide information related to the patient's ventilation region, i.e. Overdistension, Linear or Atelectasis.

In this case, the nonlinear state equation that the estimation algorithm will use is:
\begin{equation}\label{eq:stateNonLineal}
    \dot{V}= -\frac{1}{R_{aw}} (a_1V + a_2V^2) + \frac{P_v}{R_{aw}}
\end{equation}
and the three parameter vector to be estimated is:
\begin{equation}\label{eq:nonlinear_parameter}
    \theta_{nl}=[a_1, a_2, R_{aw}]
\end{equation}

\section{Nonlinear identification algorithm}\label{sec:NonlinearIDalgorithm}

In this section, the proposed systematised identification process is presented. Beginning with the estimation of the linear model, it then uses these parameters to initialise and estimate those of the nonlinear quadratic model. 
 
The algorithm core is based on a grey-box modelling approach, since the mathematical models of the respiratory system are available (Eq. \eqref{eq:Lineal} and Eq. \eqref{eq:noLineal}) \cite{LI2021111174} \cite{ESTRADAFLORES2006931}. The parameters vector values of both models are obtained using the Levenberg-Marquardt algorithm, which is an iterative method for nonlinear least squares problems, widely adopted in a broad spectrum of disciplines \cite{gavin2019levenberg} \cite{lourakis2005brief}.

The algorithm presented in this work consists of three stages:
\begin{itemize}
    \item \textbf{Signals acquisition}, where pressure, flow and volume signals are measured.
    \item \textbf{Parameter estimation}, where the signals are used to obtain the nonlinear set of parameters, one for each respiratory cycle.
    \item \textbf{Fit evaluation}, where the output of the estimated models is compared with the real output of the system. The model is saved or discarded depending on a normalised root-mean-square-error criterion.

\end{itemize}

\subsection{Signal acquisition}

The algorithm uses the pressure and flow signals measured in the mouth of a totally sedated patient. These data have been obtained with the acquisition device FluxMed\textregistered \hspace{0.5mm} GrE with a sampling frequency of 256 Hz. 

The signal pressure, $P_v$, is measured using a differential pressure sensor, which provides the variable pressure at the mouth.

On the other hand, the flow signal $F$ is measured through a fixed-orifice pneumotachograph, in which the pressure difference on both sides of the orifice is proportional to the flow value squared.

$V$ is obtained indirectly, using trapezoidal integration of the flow signal. This numerical integration is performed for each respiratory cycle.

To close this stage, an auxiliary variable $Aux$ is set equal to zero, when the end of cycle is detected. This variable will be used, during subsequent stages, to prevent a possible loop when the estimation fails.

\subsection{Parameter estimation}\label{sec:ParEst}
The proposed parameter estimation process can be summarised as follows (see the flow chart in Fig. \ref{fig:DiagFlujo} for further clarification):

\begin{figure}[]
    \centering
    \includegraphics[width=\columnwidth]{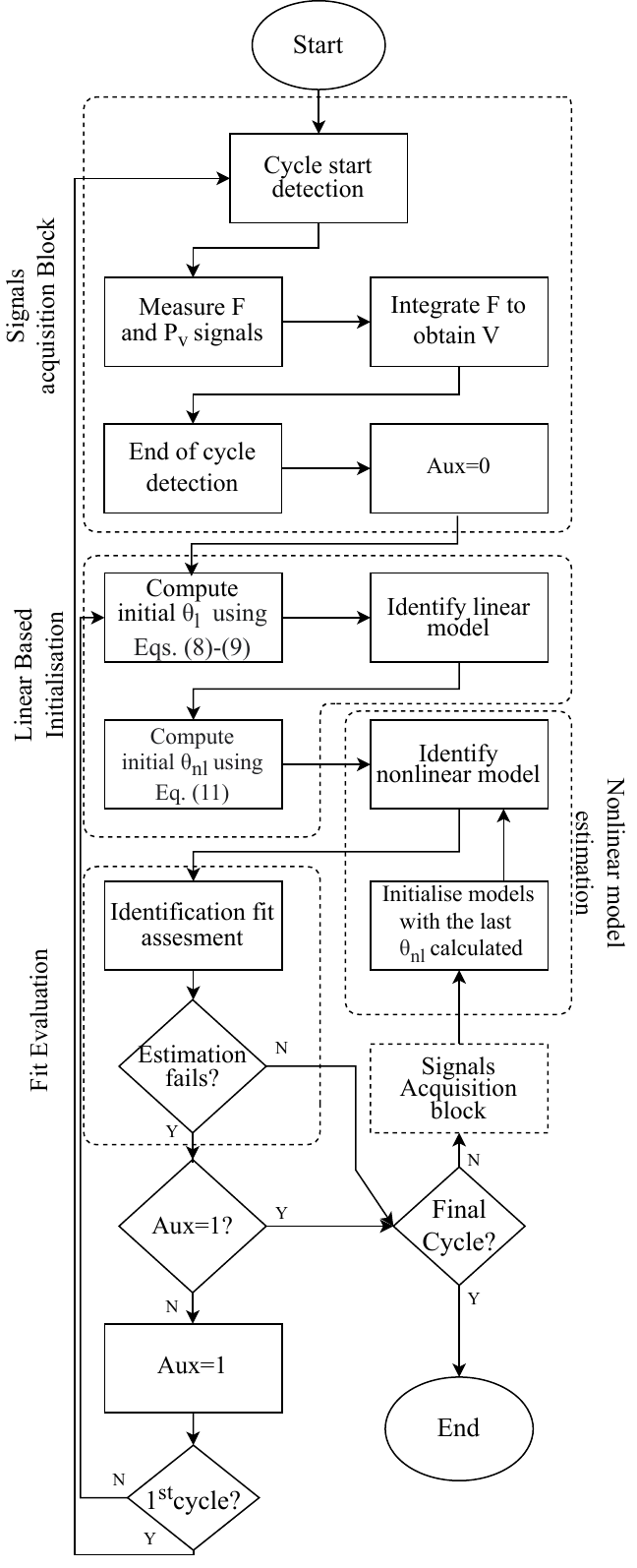}
    \caption{\label{fig:DiagFlujo} Proposed algorithm flow chart. 
    }
\end{figure}

\begin{itemize}
    \item \textit{Linear model initialisation.} The algorithm computes initial values for both linear parameter, $C$ and $R_{aw}$, using data from the first cycle of the registered signals, as indicated:
    \begin{align}
        &C=\frac{V_{Tinsp}}{P_{Plat} - PEEP} \label{eq:Cap}\\
        &R_{aw}=\frac{PIP-P_{Plat}}{PIF} \label{eq:Raw}
    \end{align}

   where $V_{Tinsp}$ is the maximum tidal air volume that enters during the inspiration semicycle, and $P_{Plat}$ is the pressure value when $F=0$ (in the change from inspiration to expiration).
    \item \textit{LM identification for NLM initialisation.} The algorithm finds the optimal values of $\theta_{l}$ for the linear model through identification, aiming to minimise the quadratic cost function:
    \begin{equation}\label{eq:MinCuad}
        M_N(\theta_l)=\sum_{k=1}^{n}(\hat{V}-V)^2
    \end{equation}
    where $n$ is the number of samples of the respiratory cycle and $\hat{V}=f(P_v,\theta_l)$ is the tidal volume computed by the model (at this step, the linear one) with $P_v$ as input signal.
    \item \textit{Nonlinear quadratic model initialisation.} The initial values for the nonlinear model identification are set based on the linear parameters estimated in the above step. So, the previous value of $R_{aw}$ is directly utilised for initialisation, while the initial values of parameters $a_1$ and $a_2$ are obtained from Eq. \eqref{eq:stateNonLineal}, which can be rewritten as:
    \begin{equation}
        P_v-FR_{aw}=(a_1V + a_2V^2).
    \end{equation}
    Then, the initial values of the nonlinear description of $P_c$, $a_1$ and $a_2$, are computed by least squares adjustment.
    
     \item \textit{Nonlinear quadratic model estimation.} With the $\theta_{nl}$ vector properly initialised, the algorithm starts an identification process to obtain the nonlinear model (Eq. \eqref{eq:MinCuad} is minimised using the output of the NLM, $\hat{V}=f(P_{B_v},\theta_{nl})$).
    
    For subsequent cycles, the algorithm initialises the identification process with the parameter values of the NLM computed in the previous respiratory cycle. 
    
\end{itemize}

\subsection{Fit evaluation}

In this stage, the estimated nonlinear model for each respiratory cycle is validated using the normalised root-mean-square-error ($NRMSE_{\%}$) criterion:
\begin{equation}
    NRMSE_{\%}=100\left(1-\frac{||V-\hat{V}||}{||V-\bar{V}||}\right)
\end{equation}
where $\bar{V}$ is the mean value of $V$. 
In this work, the validation stage uses a threshold to decide whether to keep or discard the estimated model.

If the result is higher than this threshold, the algorithm saves the estimated nonlinear model parameters. Then, it executes the Signal Acquisition Block, subsequently initialising the next nonlinear model estimation. Note that different approaches can be used to design the threshold. In the present case, the threshold was set equal to the $NMRSE_{\%}$ obtained from the standard linear model as computed by the ventilator (according to \eqref{eq:Cap} - \eqref{eq:Raw}).

On the other hand, if the result is lower than the threshold, the estimation process was unsuccessful. In this case, if $Aux=0$, the algorithm goes to Linear Based Initialisation for a second chance with different initial condition. By contrast, if such second chance fails ($Aux=1$), the algorithm discards the current respiratory cycle and continues with the next one, by the Signal Acquisition Block. 

In the specific case that the estimation fails for the first cycle, the algorithm discards it instead of returning to Linear Based Initialisation.

\section{Results from simulated and real patients}\label{sec:Results}

This section presents results obtained with simulated and  real patients data. The first subsection is intended to validate the effectiveness of the proposed algorithm for the estimation of pulmonary system parameters. To this end, the nonlinear quadratic identification algorithm is thoroughly assessed using  multiple signals generated by simulation and representative results are displayed. Once the suitability of the estimation algorithm is proved, it is used, in the second subsection, to characterise actual pulmonary parameters computed using data obtained from real sedated patients with COVID-19 under different conditions of mechanical ventilation.

\subsection{Algorithm validation with simulated patients}

Patients are simulated to obtain airflow and pressure signals to evaluate, analyse and validate the algorithm. The main objectives are, firstly, to confirm the superiority of the quadratic (nonlinear) identification over the linear one, to locally approximate a simulated patient characterised with a sigmoidal model, particularly in the Atelectasis and Overdistension Regions.
Secondly, to draw conclusions regarding the degree of information that the estimator can provide when dealing with models with hysteresis, which render a more accurate representation of real patients.

\vspace{3pt}

$\boldsymbol{(a)}$ For the first objective, the proposed identification algorithm is tested using data acquired from a sigmoidal-modelled patient without hysteresis, i.e., the P-V curve of the simulated patient was programmed according to the sigmoid curve equation \cite{venegas1998comprehensive}:
\begin{equation}
    V = a + \frac{b}{1+e^{-(P_A - c)/d}}
\end{equation}
where a, b, c, d are constants, and $V$ goes from 0 to TLC-FRC.

\begin{figure}[ht]
    \centering
    \includegraphics[width=\columnwidth]{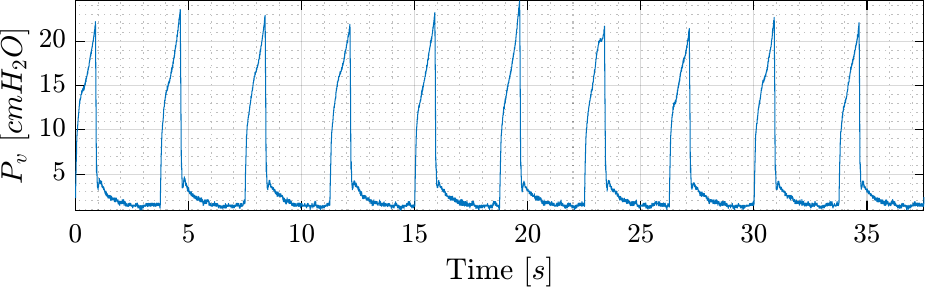}
    \caption{\label{fig:InputSignal} Pressure from actual sedated patient, used as input to the simulator.}
\end{figure}

As input signal for the simulation trials, actual pressure measured from a sedated patient was utilised (see 10 illustrative $P_v$ cycles in Fig. \ref{fig:InputSignal}). Simulations were carried out to obtain the corresponding flow signal $F$ at different PEEP values, near the LIP with PEEP=4 $cmH_2O$, in the Linear region with PEEP=13 $cmH_2O$ and near the UIP with PEEP=22 $cmH_2O$. 

\begin{figure}[ht]
    \centering
    \includegraphics[width=\columnwidth]{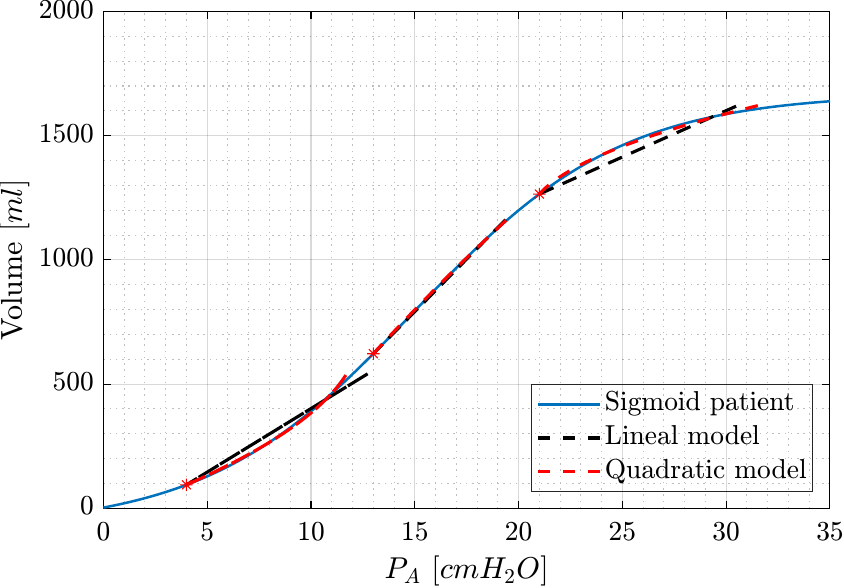}
    \caption{\label{fig:DataVsModelos} P-V curve of the sigmoid-modelled simulated patient (blue). The $P_c(V)$ estimated by the algorithm (LM in black, NLM in red) are also shown.}
\end{figure}

The algorithm was run for every simulation, so that linear and nonlinear estimated models were obtained. Figure \ref{fig:DataVsModelos} shows the $P_c(V)$ of both LM in black, NLM in red, superimposed on the P-V curve of the simulated patient (in blue). It can be appreciated that the quadratic model fits better than the linear model when the patient is ventilated in LIP ($94.40\%$ vs $99.01\%$) and UIP region ($89.10\%$ vs $97.28\%$). As expected, for the patient ventilated in the linear region, both models achieve similar good fit ($99.28\%$ LM, $99.60\%$ NLM).

Subsequently, to further assess the proposed algorithm under non-ideal conditions, uniformly distributed noise was added to both $P_v$ and $F$ signals, to account for measurement and digitalisation errors introduced by the sensors and the ADCs of the FluxMed\textregistered \hspace{0.5mm} GrE \cite{widrow1996statistical}. For these tests, simulations were conducted considering three patients, each one modelled with different sigmoidal P-V curves and $R_{aw}$ values and selected PEEP values were used to cover the whole ventilation range.

\begin{figure}[ht]
   \centering
    \includegraphics[width=\columnwidth]{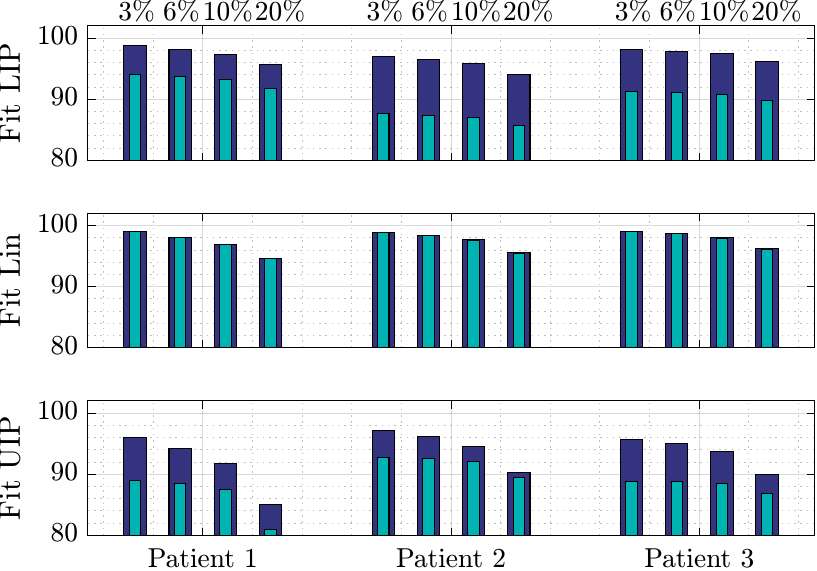}
   \caption{\label{fig:Barras} Fit results obtained by the algorithm using input signals with different percentages of added noise (column indicators at the top)}. The simulated patients were ventilated in LIP, linear and UIP regions. Bars in blue correspond to the NLM and green bars to the LM.
\end{figure}

The fits obtained with the estimated models are presented in Fig. \ref{fig:Barras}, where the blue bars correspond to the NLM, and the green ones to the LM. It can be seen that even with high levels of noise, the estimated quadratic models have really good and higher fit values than the linear ones in the LIP and UIP Regions, while in Linear Region both models have similar good results.

Note that the noise considered in this tests  significantly exceeds the measurement and digitalisation errors bounds provided by the manufacturer of the FluxMed\textregistered \hspace{0.5mm} GrE, which are $3\%$ for the $P_v$ signal and $3.5\%$ for the $F$ signal. Consequently, these results confirm the proficiency of the proposed nonlinear quadratic based algorithm for the identification in the different regions of sigmoidal-modelled patients, even under non-ideal conditions.

\vspace{3pt}
$\boldsymbol{(b)}$ Next, the second objective of evaluating the algorithm capability to draw information from more realistic data is addressed. To that effect, the  nonlinear quadratic identification algorithm was tested with data obtained from 10 simulations of a patient with nonlinear-modelled lungs that present hysteresis in the $P_{c}(V)$ curve, using the two ventilation modes (see section \ref{ssec:ventilation}).

\begin{figure}[htb]%
    \centering
        \subfigure[Simulated Patient ventilated in volume control mode.]{\includegraphics[width=\columnwidth]{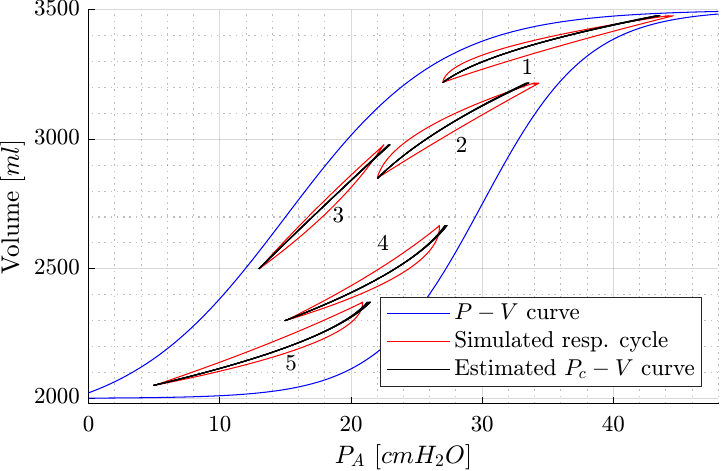}}
        \subfigure[Simulated Patient ventilated in pressure control mode.]{\includegraphics[width=\columnwidth]{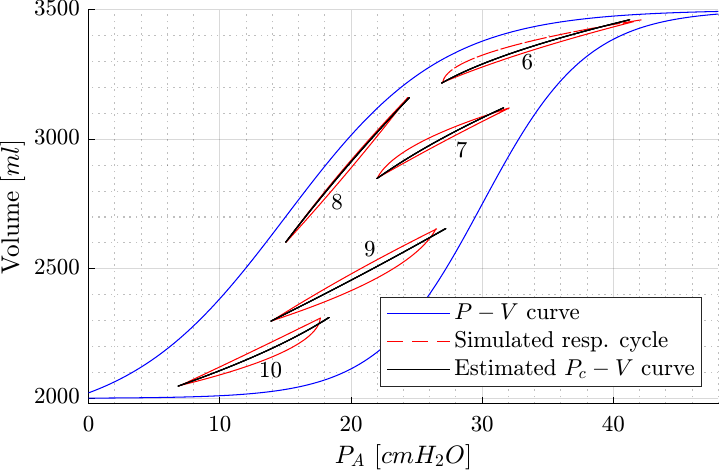}}
    \caption{In blue, P-V curve programmed in the simulator. The simulated respiratory cycles (dashed red) can be compared with the $P_c-V$ curves estimated by the NLM of the proposed algorithm (in black).} \label{fig:Hist_curva}
\end{figure}

Several breathing zones were considered, covering the range from FRC to TLC, with representative hysteresis curves for each of the three respiratory regions. The aforementioned pressure $P_v$ from an actual sedated patient (Fig. \ref{fig:InputSignal}) was also used as input signal in each simulation, to obtain the corresponding flow signals. The simulated curves are presented in dashed red lines in Fig. \ref{fig:Hist_curva}. The hysteretic characteristics were constructed based on the descriptions provided in reference  \cite{muniz2009pressure}. For their design, it was ensured that both $V$ and $P_c$ fell within actual acceptable breathing ranges.

Then, the corresponding pairs $P_v$-$F$ were fed to the algorithm, and the estimation process was conducted. The resultant estimated $P_{c}(V)$ curves are depicted in black in Fig. \ref{fig:Hist_curva}, where it is worth noting that they are mostly confined inside their corresponding modelled hysteresis curve, and their corresponding estimated parameters are shown in Table \ref{tab:resultsSim}.

\begin{table}[ht]
\centering
\begin{tabular}{|c|cccc|}
\hline

Sim \# & $a_{1}$ & $a_{2}$  & $a_{1}V_{T}$ & $a_{2}V_{T}^{2}$ \\ \hline
1      & 0.026   & 1.45e-4  & 6.7         & 9.7             \\
2      & 0.025   & 1.84e-5  & 7.4         & 4.1             \\
3      & 0.020   & 1.38e-6  & 9.6         & 0.3             \\
4      & 0.052   & -5.09e-5 & 19.1        & -6.8            \\
5      & 0.084   & -1.02e-4 & 26.9        & -10.5           \\
6      & 0.036   & 9.7e-5   & 8.6         & 5.6             \\
7      & 0.028   & 2.80e-5  & 7.6         & 2.1             \\
8      & 0.015   & 3.36e-6  & 8.4         & 1.1             \\
9      & 0.040   & -7.29e-6 & 14.1        & -0.8            \\
10     & 0.056   & -4.83e-5 & 14.6        & -3.3            \\ \hline
\end{tabular}
\caption{Identification results: second and third columns show the $P_c(V)$ estimated parameters (in $cmH_2O/ml$ and $cmH_2O/ml^2$, respectively). Fourth and fifth columns show the ranges for the linear and quadratic pressure terms of $P_c$, respectively (in $cmH_2O$).}
\label{tab:resultsSim}
\end{table}
 
It can be observed that the concavity of the estimated curves has a direct relation with the respiratory region where the patient is ventilated. For instance, when the estimated parameter $a_2$ is negligible, it means that the patient is ventilated in the Linear Region, while $a_1$ gives information about the pulmonary compliance. On the other hand, the more negative $a_2$, the more into the Atelectasis Region is the patient breathing. Conversely, the more positive $a_2$, the higher the risk of alveolar overdistension. Moreover, the physicians could use the estimated parameters together with empirical information of the patient (for instance, from  titration manoeuvre data) to have a better knowledge of their condition. In this way, they could establish quantitative ranges for $a_1$ and $a_2$ that more precisely indicate the respiratory region where the ventilation is conducted. It seems likely that further analysis could provide a relation between those parameters and the patient's Stress Index, whose value quantitatively represents whether or not the patient is being ventilated in the linear region \cite{ranieri2000pressure}.

With the above satisfactory results at hand, the proposed nonlinear identification algorithm proves to be a powerful tool to extract information from the ventilated patient. Therefore, it will be used next with real patients data.

\subsection{Algorithm applied to actual patients data}

In this section, the algorithm is tested with input-output signals ($P$-$F$) from sedated patients under assisted ventilation, measured during a PEEP titration manoeuvre.

Such manoeuvre is intended to recruit alveoli, aiming to ventilate the patient in the linear region, where both the alveolar collapse and the overdistension are minimum. In this way, the pulmonary homogenisation would be maintained, protecting the lungs  from mechanical ventilation injuries \cite{hess2015recruitment} \cite{costa2012bedside} \cite{borges2006reversibility}. It consists of increasing the PEEP value in stages until reaching a maximum admissible pressure and then decreasing it, enabling to detect the highest compliance region in the process.

\begin{figure}[ht]%
    \centering
    \includegraphics[width=\columnwidth]{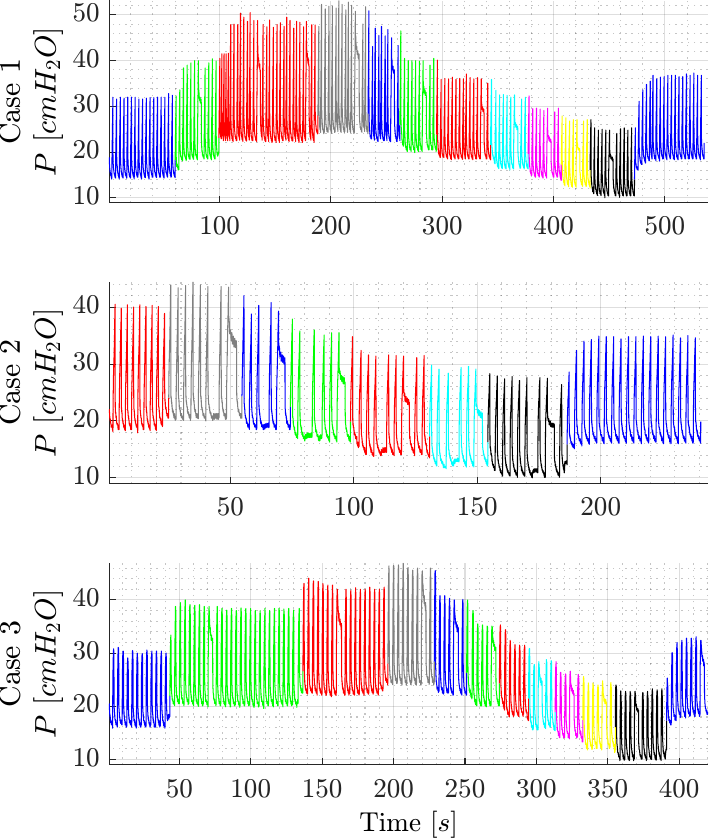}
    \caption{$P$ signal measured on two patients during PEEP titration manoeuvres. Cases 1 and 2 correspond to the same patient, two days apart.}
    \label{fig:SeñalesCOVID}
\end{figure}

\begin{figure}[]%
    \centering
        \subfigure[Results using the linear model.\label{fig:CurvasPRSa}]{\includegraphics[width=\columnwidth]{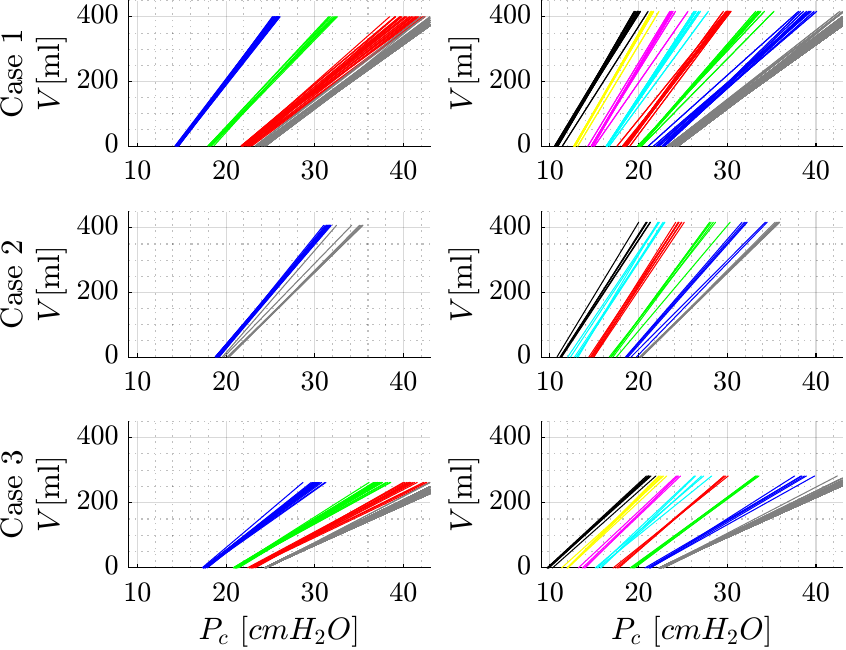}}
        \subfigure[Results using the quadratic model (NLM).\label{fig:CurvasPRSb}]{\includegraphics[width=\columnwidth]{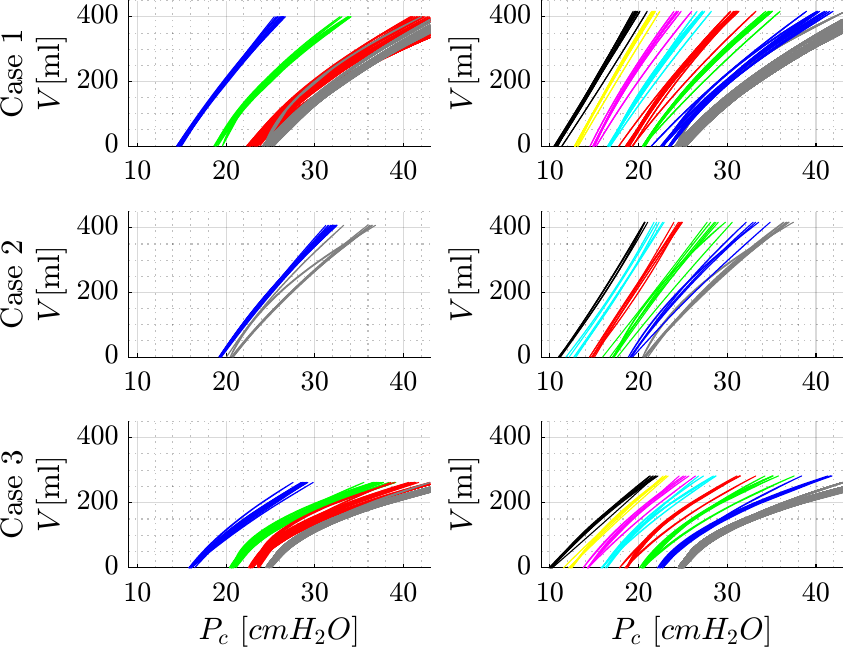}}
    \caption{$P_{c}-V$ curves identified by the algorithm in each respiratory cycle of patients measured during PEEP titration manoeuvres. Cases 1 and 2 correspond to Patient \#1 and Case 3 to Patient \#2. The left-hand side column shows the curves estimated with the ascendant PEEP series, and the right-hand side the curves estimated with the descendant PEEP series.} \label{fig:CurvasPRS}
\end{figure}

Three case studies are considered, shown in Fig. \ref{fig:SeñalesCOVID}. In the first two cases, the $P$ signals are measurements from the same patient, under PEEP titration manoeuvres conducted two days apart. Case 3 corresponds to a different patient, also under a PEEP titration recruitment manoeuvre. Note that to facilitate results interpretation, particularly to visualise the correlation between Fig.\ref{fig:SeñalesCOVID} and Fig.\ref{fig:CurvasPRS}, the signals were distinguished by coloured sections, where grey corresponds to the highest level of PEEP in the manoeuvre, and black to the lowest one.

\begin{figure}[ht]%
    \centering
    \includegraphics[width=\columnwidth]{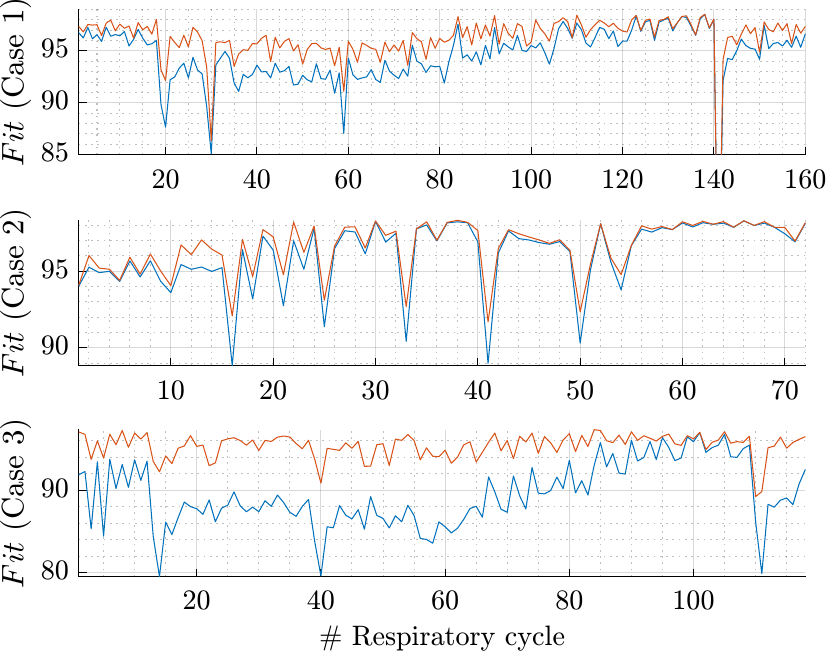}
    \caption{Fit results obtained with the linear model (in blue) and the nonlinear model (in red), using the NRMSE criterion. }
    \label{fig:Ajustes}
\end{figure}

Fig. \ref{fig:Ajustes} comparatively displays the fit obtained with both identification models (LM in blue and NLM in red). For the three cases under study, the NLM demonstrates a better fit than the LM when the patient is ventilated with high PEEP levels (corresponding to the Overdistension Region or its immediate environs). Conversely, when the PEEP decreases and the patient is ventilated in the Linear Region, the fits of both models are similar.

In this context, Case 3 is particularly illustrative. It can be appreciated that a considerable worse fit is obtained with the LM. From this information, it can be inferred that the PEEP titration manoeuvre was conducted mainly in the Overdistension Region, rather than in the Linear one. 

The P-V curves results provided by the proposed estimation algorithm are depicted in Fig. \ref{fig:CurvasPRS}, subfigure (b), while in subfigure (a) the results using the linear model of the respiratory system are shown. For the sake of clarity, the curves obtained with the ascendant PEEP series are presented in the left-hand side column and those obtained with the descendant PEEP series are in the right-hand side column.

Note that the slopes of the lines obtained with the linear model (Fig. \ref{fig:CurvasPRSa}) give information about an ``average" compliance for the different regions where the patient is ventilated, it being expected that the highest corresponds to the Linear Region. This "average" compliance is typically used by the physicians, but it is incomplete if compared to the information available from the proposed quadratic identification algorithm. In effect, the slopes of the linear identification on their own, in certain cases, do not suffice to determine if the patient is outside the Linear Region. The information regarding curvature is of decisive importance.

The best way to illustrate this situation is with Case 3. A standard PEEP titration manoeuvre was performed on Patient \#2. As a result, a series of straight lines were obtained with the LM (Fig. \ref{fig:CurvasPRSa}  bottom). An analysis of their slopes (relatively low, if it is assumed the titration has covered a great deal of the Linear Region) might lead to the conclusion that the patient suffers from a severe reduction of compliance. The actual situation is quite different and it can be easily detected from the curves in Fig. \ref{fig:CurvasPRSb}, Case 3. As it was previously mentioned, the curvatures reveal that the titration manoeuvre was conducted mainly in the
Overdistension Region.

The following provides a more detailed analysis of the results obtained with the proposed nonlinear identification algorithm. Focusing the attention in Fig \ref{fig:CurvasPRSb}, in Cases 1 and 2 (Patient\#1), right column,  a slight change of concavity can be observed (concave left, i.e. negative $a_2$, for the lowest values of PEEP). This means that the patient is entering into the Atelectasis Region. Therefore, it is not desirable to go further below that minimum PEEP with this patient. On the other hand, in Case 3 (Patient\#2) the concavity has not yet changed ($a_2$ is  positive) for the lowest PEEP value. With this result, the physician can infer that there is still room for PEEP reduction in Patient\#2's titration manoeuvre.

For its part, information of the top end of the PEEP titration manoeuvre can be easily obtained from the left-hand side column of Fig. \ref{fig:CurvasPRSb}. For instance, the pronounced curvature of Patient\#2 during the highest PEEP is a clear indicator that the patient is ventilated in the Overdistension Region. Consequently, the maximum PEEP applied during its titration manoeuvre is possibly excessive. This is the kind of online feedback that clinicians can use to make the decision whether or not to increase the PEEP, since in this case the patient  could generate a lung injury due to the overdistension of alveolar units.

The aforementioned is not the only information about the lungs condition that can be extracted from the algorithm results. The slopes of the curves in Fig. \ref{fig:CurvasPRSb} provide an insight on the patient's compliance. Moreover, they are inversely related to parameter $a_1$, corresponding this parameter to the lung's elastance in the Linear Region.  

Therefore, the steeper the slope (lower $a_1$), the higher the compliance. For instance, comparing  Patient\#1 against Patient\#2, it is apparent that the lung of the latter presents a lower compliance (or expandability), requiring more pressure to move the same amount of tidal volume. 

\begin{figure}[htb]%
    \centering
    \includegraphics[width=\columnwidth]{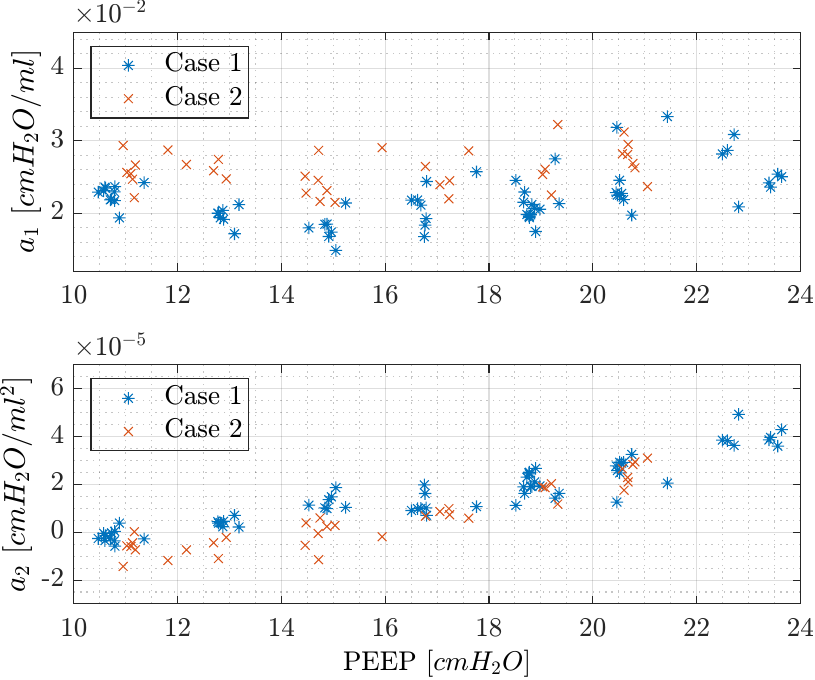}
    \caption{Comparison of parameters $a_1$ (above) and $a_2$ (below) estimated for Cases $1$ and $2$, according to the corresponding PEEP value.}
    \label{fig:Comp_Pat1}
\end{figure}

Such comparative analysis can also help to assess the evolution of a single patient over time. For this example, let's compare Case 1 and Case 2, recalling that those results were obtained from Patient\#1's data collected with a difference of two days. Examining the curves in Fig. \ref{fig:CurvasPRSb}, it can be appreciated that the patient's condition is relatively stable. However, from direct observation of the slopes, hints of minor decline seems to be manifesting. This can be better confirmed by plotting parameters $a_1$ and $a_2$ against the PEEP in Fig. \ref{fig:Comp_Pat1}. Effectively, a growing trend in $a_1$ is discernible, implying a compliance reduction. Additionally, comparative inspection of parameter $a_2$ shows that, two days later, Patient\#1 is approaching the Atelectasis Region with higher values of PEEP  (note that in Case 2, the sign of $a_2$ is negative for PEEPs below 15 $cmH_2O$).

Of course, for conclusive medical actions, more complementary data and clinical evidence need to be taken into consideration, but these examples illustrate how the information provided by the proposed algorithm may assist health professionals with decision-making tasks, treatment and diagnosis.

As an additional remark, comparing the employed techniques, based on the Levenberg-Marquardt algorithm, with the well-known least square fitting (LSF) techniques, our approach achieves higher fit to the data, but at the expense of more processing time. Depending on the microprocessor and programming, this extra time may cause that one cycle every other is not identified. This is not significant, as it is not an uncommon loss in respiratory monitoring.

\section{Conclusions}\label{sec:Conclusions}

An algorithm capable of estimating, online and offline, the internal parameters of the pulmonary system of patients under assisted ventilation has been presented. With noninvasive measurements of pressure and flow at the patient's mouth and nonlinear identification techniques, the algorithm estimates the parameters of the quadratic model that best fits the patient in each respiratory cycle.

Firstly, the performance of the nonlinear identification algorithm was analysed and validated via simulation. At this stage, the proposed hierarchical identification algorithm was tested using a nonlinear sigmoidal modelled patient. Results show that both approaches, the LM and the NLM, provide excellent fits in the Linear Region. However, an important improvement of the fit is attained with the proposed NLM based estimation method when the patient was ventilated outside the linear region. Furthermore, to account for even more stringent and realistic conditions, the algorithm was subsequently tested considering noise, resulting in a remarkably  better performance of the NLM approach over the LM in the Atelectasis and Overdistension regions.

Once the algorithm proficiency was established, hysteresis was included in the simulation models attaining a more realistic characterisation of the patients. From those tests, it was concluded that the parameters estimated by the algorithm, $R_{aw}$, $a_1$ and $a_2$, can provide very useful online information about the airflow resistance, the lung's compliance and the region where the patient is currently ventilated, respectively. 

Boosted by those successful results, in a final stage, the proposed algorithm was used to characterise actual sedated patients with COVID-19 under assisted ventilation, during PEEP titration manoeuvres. Three case studies were considered, where two of them correspond to the same patient, in different days, and the last one to another patient. From the concavity of the curves obtained for each respiratory cycle, quantified by parameter $a_2$, the algorithm proved to be a useful tool to determine when the patient moves out from the Linear Region, into the Atelectasis or the Overdistension Regions. Also, the slope of the curve and its evolution, quantified by parameter $a_1$, supply information to the clinician about the thoracic-pulmonary compliance and its progress over time.

As future work, it is planned the use of the proposed estimation tool on a large number of patients, to correlate pathologies, health evolution and other representative data, which will allow for comparison and a comprehensive study. It is expected that this will provide quantitative indicators to better characterise the patient's status and the ventilation conditions, helping the physician to better diagnose, to adjust therapy and to improve recovery.

\section*{Competing Interest}
The authors have no relevant financial or nonfinancial competing interests to disclose.

\section*{Founding}
This work was carried out with the support of the Facultad de Ingenier\'{i}a, Universidad Nacional de La Plata (UNLP project 11-I255), CONICET (PIP 112-2020-010281CO and PIP 112-2020-010281CO), I + D + i Agency (PICT N° 2018-03747), Bunge y Born Foundation - I + D + i Agency (Project COVID-19 \# 873) and Hospital I.E.A y C. San Juan de Dios de La Plata.

\end{document}

%% file: astro.tex
\def\point#1{\hbox{\setbox7=\hbox to0.6em{\hfil.\hfil}%
\setbox8=\hbox to0.5em{\hfil$^{#1}$\hfil}%
\box7\kern-0.5em\box8}}

\def\pointmin#1{\hbox{\setbox2=\hbox to0.8em{\hfil.\hfil}%
\setbox3=\hbox to0.6em{\hfil$^{#1}$\hfil}%
\box2\kern-.7em\box3}}

\def\mmm{\pointmin{\mathrm{m}}\kern.15em}

